\documentclass{pazh6286}
\usepackage{graphicx}
\usepackage{latexsym}

\newcommand{\km}{\,\mbox{km}\,\mbox{s}^{-1}}

\def\farcm{\hbox{$.\mkern-4mu^\prime$}}
\def\farcs{\hbox{$.\!\!^{\prime\prime}$}}

\begin{document}

\title{The phenomenon of the galaxy NGC 6286:
a forming polar ring or a superwind? }

\author{
L.V. Shalyapina$^1$ \and A.V. Moiseev$^2$ \and V.A.Yakovleva$^1$ \and
V.A. Hagen-Thorn$^1$  \and A. N. Burenkov$^2$}

\institute{ $^1$ Astronomical Institute, St. Petersburg State University,
Universitetsky pr.28, Petrodvorets, 198504 Russia
\\
$^2$ Special Astrophysical Observatory, Russian Academy of Sciences, Nizhnii
Arkhyz, Karachai-Cherkessian Republic, 357147 Russia
 }

\offprints{L.V.~Shalyapina, \email{lshal@astro.spbu.ru}}
\date{Received March 20, 2003 }

\titlerunning{The phenomenon of the galaxy NGC 6286... }
\authorrunning{ Shalyapina et al.}

\abstract{ We present the observations of the pair of interacting
 galaxies NGC6285/86 carried out with the 6m Special
Astrophysical Observatory (SAO) telescope by using 1D and 2D spectroscopy. The
observations of NGC6286 with a long-slit spectrograph (UAGS) near the
$H_\alpha$ line revealed the rotation of the gaseous disk around an axis
offset by $5''-7''$ from the photometric center and a luminous gas at a
distance up to 9 kpc in a direction perpendicular to the galactic plane. Using
a multipupil fiber spectrograph (MPFS), we constructed the velocity fields of
the stellar and gaseous components in the central region of this galaxy, which
proved   to be similar. The similar line-of-sight velocities of the pair and
the wide ($5'\times 5'$) field of view of the scanning Fabry-Perot
interferometer (IFP) allowed us to obtain images in the $H_\alpha$ and [NII]
$\lambda6583$ emission lines and in the continuum as well as to construct the
line-of-sight velocity  fields and to map the distribution of the
[NII]$\lambda6583/H_\alpha$ ratio for both galaxies simultaneously. Based on
these data, we studied the gas kinematics in the galaxies, constructed their
rotation curves, and estimated their masses ($2\cdot 10^{11}M_\odot$ for
NGC6286 and $1.2\cdot 10^{10}\,M_\odot$ for NGC6285). We found no evidence of
gas rotation around the major axis of NGC6286, which argues against the
assumption that this galaxy has a forming polar ring. The IFP observations
revealed an emission nebula around this galaxy with a structure characteristic
for superwind galaxies. The large [NII]$\lambda6583/H_\alpha$ ratio, which
suggests the collisional excitation of its emission, and the high infrared
luminosity are additional arguments for the hypothesis of a superwind in the
galaxy NGC~6286. A close encounter of the two galaxies was probably
responsible for the starburst and the bipolar outflow of hot gas from the
central region of the disk}

\maketitle

\centerline{\large \bf INTRODUCTION}

\medskip

 NGC6285/86 (Arp 293) is a pair of interacting galaxies
(Fig. 1) with similar luminosities and radial velocities\footnote{According to
NED data}. They are $\sim$1\farcm5 apart, which corresponds to $\sim40$ kpc
for a distance $D = 91$ Mpc (at $H_0 = 65 \km\, Mpc$ and $V_{sys}^{gal} = 5925
\km$; see below). A faint `bridge' can be seen between the galaxies in deep
CCD images. Whitmore et al. (1990) included one of the galaxies from the pair,
NGC6286, in the catalog of polar-ring galaxies as a possible candidate (C51),
because a diffuse structure (semi-ring) located to the SE of the main body of
the galaxy is clearly seen in the reproduction in the atlas by Arp (1966).

 NGC6286 is a spiral Sb-type (RC3) galaxy seen almost edge-on with
a thick dust lane that runs inclined  to the stellar disk of the
galaxy. Since, according to IRAS data, this galaxy has a high
infrared luminosity, $\log(L_{FIR}/L_B) = 11.28$ (Soifer et al.
1987), it was included in the studies of bright infrared galaxies
(see, e.g., Baan 1989; Young et al. 1989). The infrared fluxes from
NGC6286 were measured at various wavelengths (Soifer et al. 1987,
1989). These fluxes were used to determine the dust temperature and
mass (Young et al. 1989) and to estimate the star-formation rate,
$SFR= 56\,M_\odot\, yr^{-1}$ (Smith et al. 1998). Based on radio
data, Sanders et al. (1986) estimated the H2 mass for this galaxy,
$\log M(H_2)/M_\odot = 9.97$.

The optical spectra of this galaxy were described by several authors.
Reshetnikov \& Combes (1994) classified its nuclear spectrum as H II and
pointed out a peculiarity in the radial velocity distribution of the gas
along the minor axis of the galaxy. Veilleux et al. (1995) classified the
nuclear spectra of NGC6286 and NGC6285 as LINER and H II, respectively.
Smith et al. (1996) used the $H_\alpha$ line measurements to construct
the rotation curve for NGC6286 and estimated its mass ($M = 1.3 \cdot
10^{11}M_\odot$). Reshetnikov et al. (1996) performed a detailed
photometric study of NGC6286 in the $B,V,R_C$ bands. Based on
peculiarities of the line-of-sight velocity curve along its minor axis
and on photometric data, they assumed that the diffuse structure
(semi-ring) located at the SE edge of the galaxy is a forming polar ring.
To analyze the kinematics of the gaseous and stellar components of the
galaxy NGC6286 in more detail, we have undertaken its study by using
long-slit (1D) and 2D (panoramic) spectroscopy.

\begin{figure}

\centerline{\includegraphics[width=8 cm]{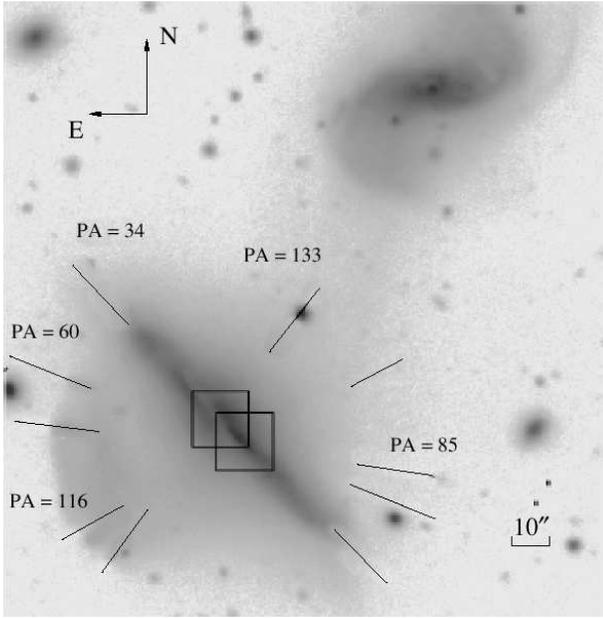}}
 \caption{
 Fig. 1. The  $R_C$-band image of the NGC6285/86 pair obtained with
the 6-m SAO telescope (for the assumed distance to the galaxies of $D
= 91$ Mpc; $1''$ corresponds to 0.44 kpc).}
\end{figure}

\section{OBSERVATIONS AND DATA REDUCTION}

The spectroscopic observations of NGC6286 were performed at the prime
focus of the 6-m SAO telescope. The first observations were carried
out with the long-slit spectrograph UAGS (Afanasiev et al. 1995) in
1997 and 1999. The observations with the multipupil fiber spectrograph
MPFS (Afanasiev et al. 2001; a description of the spectrograph can be found on the
Internet at the SAO WWW
page\footnote{http://www.sao.ru/hq/lsfvo/devices.html}) were performed in 2001, and the data with a
scanning Fabry-Perot interferometer (IFP) were obtained in 2002. A
log of observations is given in Table 1.

In UAGS, the detectors were Electron $530 \times 580$ - pixels and
Photometrics $1024 \times 1024$-pixel CCDs in 1997 and 1999, respectively. The
UAGS slit size was $2'' \times 140''$. The observed spectral range included
the $H_\alpha$, [NII]$\lambda6548, 6583$ and [SII]$\lambda6716, 6731$ emission
lines.

The UAGS spectral data were reduced in the ESO-MIDAS environment
using the LONG subroutine. After the primary reductions, we
performed a smoothing along the slit with windows of 0\farcs8 for
the central region and $2''$ starting from a distance of $15''$ from
the center. The line-of-sight velocities were measured from the
positions of the centers of the Gaussians fitted to the emission
lines. The accuracy of these measurements was estimated from the
nightsky [OI]$\lambda6300$ line to be $\pm10 \km$. We also measured
the relative intensities and $FWHM$s of the  emission lines. The
observed $FWHM$s were corrected for the width of the instrumental
profile by using the standard relation $(FWHM)^2 = (FWHM)^2_{obs}+
(FWHM)^2_{instr}.$ According to the measurements of lines from a
calibration lamp, the $FWHM$ of the instrumental profile was 3.6 \AA.
The $H_\alpha$, [NII]$\lambda6583$, and [SII]$\lambda6716, 6731$
emission lines are most intense in the spectrum of this galaxy. These
lines were used to construct the line-of-sight velocity curves for
the ionized gas.

During our MPFS observations, we took spectra from 240 spatial elements (in
the form of square lenses) simultaneously that formed an array of $16 \times
15$ elements in the plane of the sky. The angular size of a single element was
$1''$. The detector was a Techtronix $1024 \times 1024$ - pixels CCD. The
observations were carried out in two spectral ranges. The green range included
emission lines of the gaseous component ($H_\beta$, [OIII]$\lambda 4959,
5007$) and absorption features of the stellar population of the galaxy
(MgI$\lambda5175$, FeI$\lambda5229$, FeI + CaI$\lambda5270$, etc.). The red
range contained the $H_\alpha$, [NII]$\lambda 6548, 6583$, and [SII]$\lambda
6716, 6730$ emission lines.

We reduced the observations by using the software that was developed by V.L.
Afanasiev (SAO) and that runs in the IDL environment. We constructed
two-dimensional maps of the emission lines intensities and ionized gas
line-of-sight velocity  fields by using  Gaussian fitting of the emission
lines profiles. The line-of-sight velocities were determined with an accuracy
of $10-15 \km$. The line-of-sight velocity fields of the stellar component
were constructed by means of a cross-correlation technique modified for 2D
spectroscopic data and described  by Moiseev (2001). Spectra of the star HD
148293  and the twilight sky were used as templates for cross-correlation. The
line-of-sight velocities were determined from the absorption lines with an
accuracy of $\sim10 \km$.

\begin{table*}
  \caption{ A log of observations}
\begin{tabular}{lccccc}
\hline
Instrument, year & Exposure time & PA, field & Spectral range & Reciprocal dispersion & Seeing\\
                 &     s         &           &  \AA           &   \AA/px& \\

\hline
UAGS      & $3 \times 1800$ & $34^\circ$ & 6200-7000 & 1.5 & 3\farcs2 \\
Sep. 1997 & $3 \times 1800$ & $116^\circ$ & 6200-7000& 1.5 &2\farcs5\\
          & $3 \times 1800$ & $133^\circ$ & 6200-7000& 1.5 &2\farcs5\\
\hline
UAGS      & $3 \times 1800$ & $60^\circ$ &  6200-7000& 1.2 &1\farcs6\\
Aug.-Sep. 1999& $3 \times 1800$ & $85^\circ$ & 6200-7000 & 1.2 &1\farcs6\\
          & $3 \times 1800$ & $85^\circ$ & 6200-7000 & 1.2 & 1\farcs5\\
          & $3 \times 1800$ & $133^\circ$ &6200-7000 & 1.2 & 1\farcs6\\
\hline
 MPFS     & $3 \times 1200$ &  Center & 4900-6200 & 1.4    & 2\farcs0 \\
Apr.-Sep. 2001 &$3 \times 900$ &Center& 5900-7200 & 1.4    & $2-2.5''$\\
\hline
 IFP      & $32 \times 300$ & &    $H_\alpha$ + [NII] &  0.9&  1\farcs5\\
 Apr.-Sep 2002 & $32 \times 180$ & &  [NII]$\lambda6583$&  0.9&  2\farcs0\\
\hline
\end{tabular}
\end{table*}

The IFP observations were carried out by using the SCORPIO multi-mode
focal reducer. The reducer is described at the SAO
WWW-page\footnote{http://www.sao.ru/hq/moisav/scorpio/scorpio.html} its
parameters in interferometric observations were given by Moiseev (2002).

For our observations, we used the interferometer in the 235th order at a
wavelength of $6562.8$ \AA. The interfringe $\Delta\lambda = 28$\AA\,
corresponded to a $\sim1270 \km$ range free from orders overlapping.
Premonochromatization was made by using narrow-band filters with $FWHM =
19-20$\AA\, centered at the  spectral range  included the $H_\alpha$ or
[NII]$\lambda6583$ emission lines of the galaxy. During an exposure, we
sequentially took 32 interferograms of the object at various IFP's gaps, the
size of the spectral channel was about 0.9 \AA\, ($\sim40 \km$). The width of
the instrumental profile was $FWHM=2.5$\AA\, ($\sim110\km$). The detector was
the same CCD as in the MPFS observations. Since the readout was performed
with  $2 \times 2$ binning, a $512\times512$-pixel image (the pixel size was
0\farcs56) was obtained in each spectral channel.

We reduced the interferometric data by using the IDL-based software developed
at the SAO (Moiseev 2002). After the preliminary reduction (a subtraction of
night-sky lines and wavelength calibration), the observational data were
collected to data cubes where each point in a $512 \times 512 $ field
contained a 32-channel spectrum. A Gaussian smoothing along a  spectral domain
with $FWHM=1.5$ channels and a two-dimensional Gaussian smoothing in a spatial
domain with $FWHM=2-3$ pixels (depending on the seeing), was made by using the
ADHOC package\footnote{The ADHOC software package was developed by J.
Boulesteix (the Marseilles Observatory) and is freely available on the
Internet.}. To construct the velocity fields and monochromatic images, we
fitted the emission line profiles by Gaussians. The measurement errors of the
line-of-sight velocities did not exceed $10 \km$ for single lines.

However, the following factor hindered our measurements of the line-of-sight
velocities and intensities of the emission lines during our observations near
$H_\alpha$. According to the long-slit observations, the range of
line-of-sight velocities in the galactic  disk is $450-500 \km$ (see below).
On the wavelength scale, this range accounted for about half of the $FWHM$ of
the narrow-band filter used to separate  the desired spectral range (see Fig.
2 top). Therefore, during the observations with the IFP670 filter centered on
the shifted $H_\alpha$ line ($\lambda_C= 6691$\AA), the [NII]$\lambda6583$
line was also seen in the wing of the filter transmission curve in the central
and SW regions of the galaxy. Thus, for the blueshifted regions, the observed
intensity of the nitrogen line increased sharply, while the intensity of the
$H_\alpha$ line decreased. Our standard procedure of the correction for the
filter transmission curve (``a spectral flat-field'') does not allow the true
intensities of the lines in the wing of the filter transmission curve to be
restored (Moiseev 2002). In addition, in the central region of the galaxy (see
Section 2.1), the $H_\alpha$ and [NII] line IFP-profiles slightly overlap,
because the $FWHMs$ of the emission lines increase. The both effects (the
increases in relative [NII] intensity and line profile width) made it
difficult to measure the line-of-sight velocities and intensities of
$H_\alpha$ in the central and SW regions of the galactic disk.

\begin{figure}
\centerline{\includegraphics[width=8  cm]{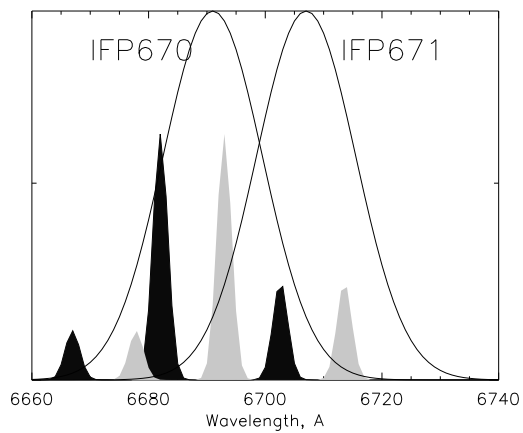}}
\centerline{\includegraphics[width=7  cm, angle=-90]{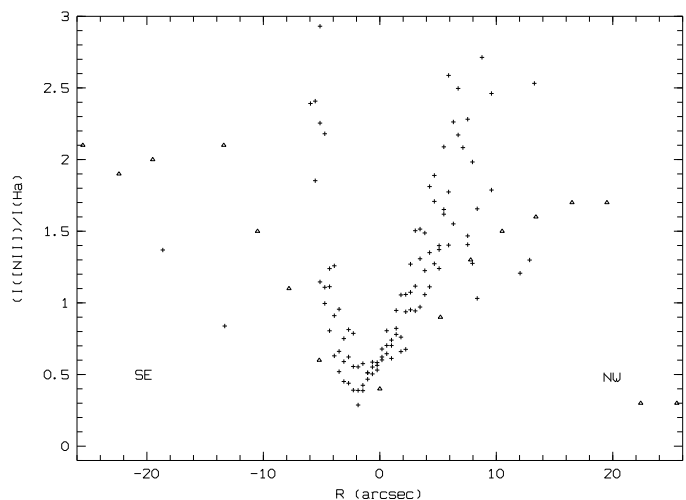}}
 \caption{
 (top) IFP
observations. The spectra of the galaxy (the $H_\alpha$  and
NII]$\lambda6548,\lambda 6583$ lines) are shown schematically. The
spectra that correspond to $V = 5450$ and $5950 \km$ are colored black and
gray, respectively. The solid lines represent the transmission curves of
both filters (IFP670 and IFP671). (bottom) The distribution of the
[NII]$\lambda6583/H_\alpha$ ratio, as constructed from UAGS (crosses) and
IFP (triangles) data (at $PA = 116^\circ$).}
\end{figure}

To improve the line-of-sight velocities in the galactic disk, we carried out
additional observations with a filter centered on the [NII]$\lambda6583$ line
(IFP671 with $\lambda_C = 6707$\AA). In this case, the filter separated  only
one spectral line (Fig. 2top).

Thus, we constructed the line-of-sight velocity fields and intensity
distributions in the $H_\alpha$ and [NII]$\lambda6583$ lines from the
observations with IFP670 and IFP671, respectively. A comparison of the
measured line intensities based on the UAGS and IFP observations shows a good
agreement (Fig. 2 bottom). We also mapped the galaxies in the continuum near
the emission lines.

\section{RESULTS}

\subsection{UAGS spectra}

We have obtained  long-slit spectra of NGC6286 near $H_\alpha$ at the five
position angles of the slit which are marked  on the  Fig.~1. The line-of-sight
velocities were measured from the $H_\alpha$, [NII]$\lambda 6548, 6583$ and
[SII]$\lambda 6716, 6730$ emission lines. The data obtained from the forbidden
lines are in close agreement, within the errors. However, since
[NII]$\lambda6583$ is more intense than the other lines, we will present the
measurements of this line only. Below, we give heliocentric velocities.

Figure 3a shows the line-of-sight velocity distribution along the major
axis ($PA = 34^\circ$). On the horizontal axis, $R = 0$ corresponds to the
position of the peak continuum intensity (the photometric center). In most
of the velocity curve, the line-of-sight velocities measured from both
lines ($H_\alpha$, [NII]) are equal, except the region $-9'' < R < 7''$
where the data obtained from these lines systematically differ by $\sim20
\km$.

The line-of-sight velocity to the SW from the photometric center is almost
constant, slightly decreasing toward the galactic edge. In the region $0
< R < 6''$, the line-of-sight velocity gradient is roughly  constant. In
the NE direction the velocity increases, but with a slightly smaller
gradient. Note that the slit crosses the dust lane here. At distances
from the photometric center larger than $22''$, the curve flattens out,
as confirmed by the data obtained later with the IFP.

Since the galaxy is seen nearly edge-on, the line-of-sight velocity curve
along its major axis may be considered to be the curve of circular
rotation (to within the line-of-sight projection of the non-circular
velocities). Within the errors, our data closely agree with the rotation
curve from Smith et al. (1996) if we take the middle point of the
line-of-sight velocity range, which is offset by about 5\farcs5 to the NE
of the photometric center, as the coordinate origin and assume that the
velocity at this point ($5650 \km$) is the velocity of the system.

The misalignment of the photometric and dynamical centers also follows
from our MPFS and IFP observations (see Sections 2.2 and 2.3) and is
probably attributable to the presence of light-absorbing dust in the
circumnuclear region. Note that there is a large uncertainty in
establishing the position of the galaxy nucleus. Since the dust lane runs
at an angle to the disk plane (Fig. 1), the isophotes shape in the
central region is distorted and asymmetric (Smith et al. 1996;
Reshetnikov et al., 1996). As a result, the intensity peak is shifted to
the SW from the nucleus, and this shift changes with wavelength due to
selective dust absorption.

The middle parts of the line-of-sight velocity curves constructed at $PA
= 60^\circ$ and $85^\circ$ have rectilinear segments. As in the previous
case, the centers of their symmetry do not coincide with the positions of
the continuum intensity peaks, and the gradients decrease with increasing
angle between the major axis of the galaxy and the spectrograph slit. The
rectilinear segments are about $5''$ long, which suggests solid-body
rotation of the gaseous galactic disk in this region. A surprising fact
is that the emission lines are observed up to a distance of 4 kpc on both
sides of the plane of the galactic disk.

The line-of-sight velocities at the spectrograph slit positions close to
the direction of the minor axis of the galaxy are differ. At $PA =
133^\circ$, the line-of-sight velocity is constant in the central region
($R< 5''-6''$), as must be the case for circular rotation of the gaseous
disk in the galactic plane. Further out as one recedes from the center on
both sides, there is a large spread in line-of-sight velocities, with the
intensity of the [NII]$\lambda6583$ line increasing compared to the
intensity of $H_\alpha$. In this case, the spectrograph slit crossed the
major axis of the galaxy at a point offset by $\sim6''$ to the NE of the
photometric center.

\begin{figure}
\centerline{\includegraphics[width=7  cm, angle=-90]{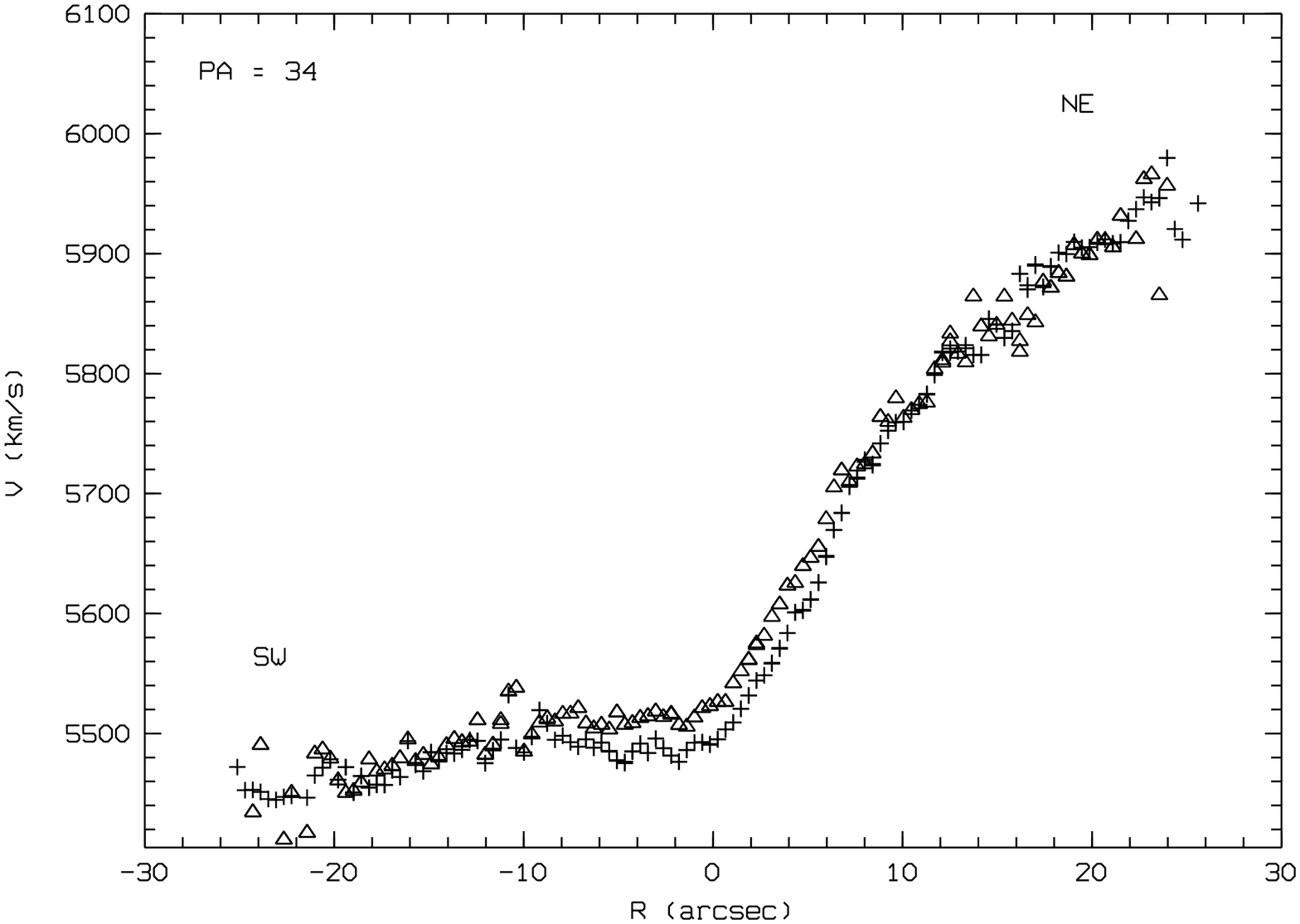}}
\centerline{\includegraphics[width=7  cm, angle=-90]{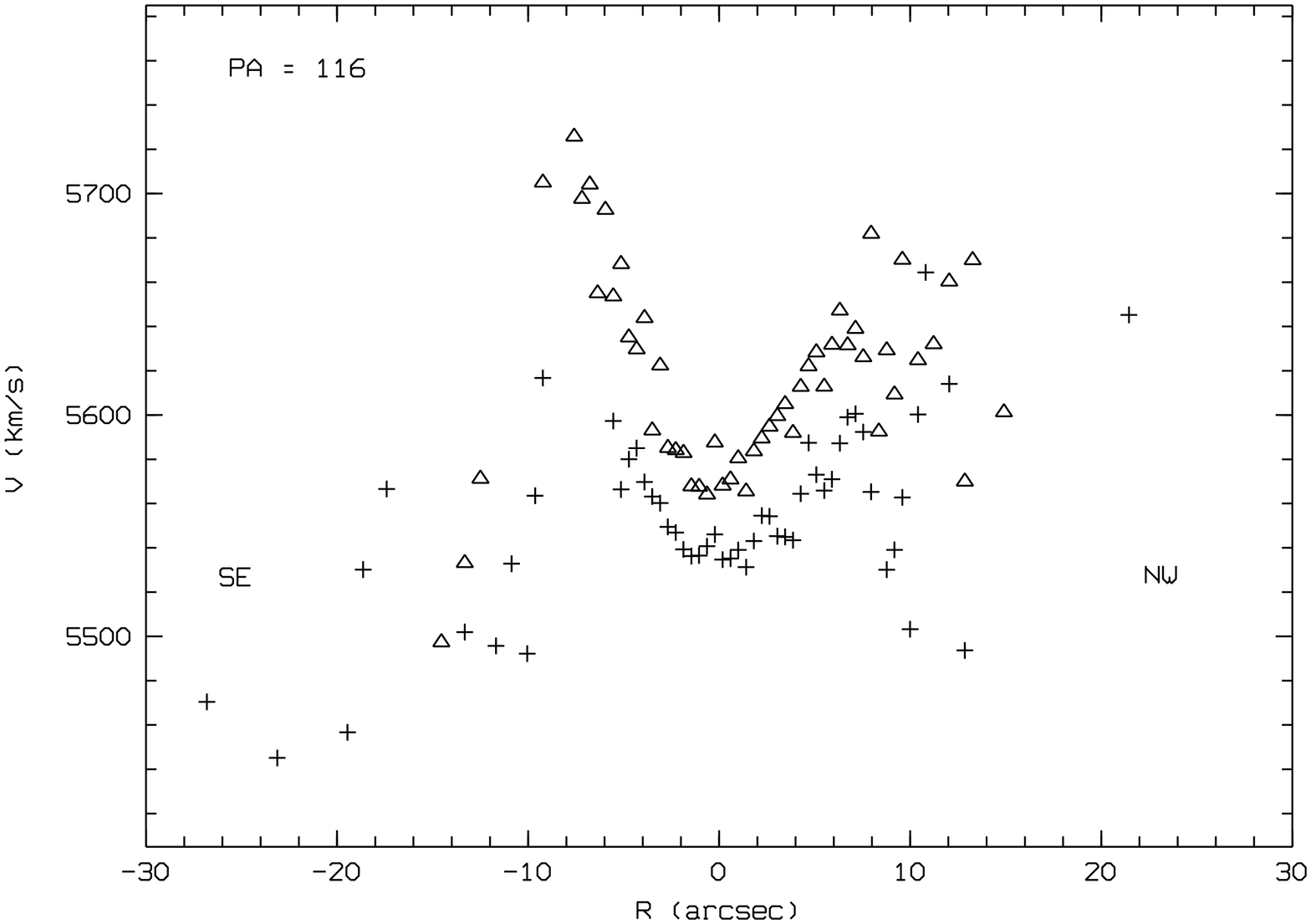}}
\caption{Line-of-sight velocity curves: (top) $PA = 34^\circ$ along the
major axis of the galaxy, (bottom) $PA = 116^\circ$; the crosses mark
$H_\alpha$ measurements, and the triangles mark  [NII]$\lambda6583$
ratio. The peak of a continuum intensity was taken as zero point on the
horizontal axis.}
\end{figure}

Let us consider the features of the line-of-sight velocity curve at $PA =
116^\circ$ shown in Fig. 3b. The emission lines are traceable far from the
disk plane, up to 9 kpc in the SE direction. In the central part of the
curve, the line-of-sight velocities increase on both sides of the center.
This behavior of the line-of-sight velocities can be explained by
assuming that the directions of the slit and the rotation axis almost
coincide and that the slit position is shifted by about $5''-7''$ to the
SW from the dynamical center. This assumption is justified in Section 2.3.

Apart from the line-of-sight velocities, we measured the relative
intensities and the $FWHMs$ of the emission lines. As we noted above, a
characteristic feature of the individual spectra is an increase in the
intensity of the [NII] line compared to the intensity of $H_\alpha$ as
one recedes from the disk plane (Fig. 2b). The [NII]/$H_\alpha$ ratio is
0.35 at the photometric center and reaches 1.5 in outer galactic regions.
Below (see Section 2.3), we will consider the distribution of relative
line intensities in more detail by using IFP data.

The UAGS data suggest the existence of a luminous gas far from the
plane of the galactic disk, but they do not allow the full picture
of its motion to be constructed. In addition, they give no
information about the stellar component of the galaxy. Such
information was obtained when NGC6286 was studied by using 2D
spectroscopy.

\subsection{ MPFS spectra}

 The MPFS observations were performed for the central
part of the galaxy. The positions of the spectrograph fields are shown in
Fig. 1. The center of the fields coincide with the photometric center in
the green spectral range and is shifted by $5''$ to the NE in the red
 range. Near and to the SW of the photometric center, the
line-of-sight velocities of the forbidden lines are higher than those of
$H_\alpha$ and $H_\beta$, as in the case of long-slit spectra. In
general, however, the velocity fields constructed from the emission lines
are similar, and we will consider the data obtained from $H_\beta$ to
compare them with the velocity field of the stellar component. The
results are shown in Fig. 4. We see that the motions of the stellar and
gaseous components are similar and generally consistent with the
assumption of circular rotation of the gas and stars, although the
isovelocities  are appreciably distorted in both fields. These
distortions are most likely attributable to the presence of dust. Since
the size of the field of view  is relatively small, we failed to
determine the position of the dynamical center accurately. We may only
note that it is shifted to the NE of the photometric center.

Based on the measured equivalent widths of absorption lines, we
determined the chemical indices $Mgb$ and $<Fe>$ in the Lick system
(Worthey et al. 1994) and constructed their radial distributions. The
indices $Mgb$ increase from 1.8 to 3, while the indices $<Fe>$  have a
small gradient and change from 1.5 to 2. The $[Mgb/<Fe>]$ ratio changes
from 0.0 dex near the photometric center to +0.3 dex at a distance of
$4''-5''$ to the NE. Using model calculations (Worthey, 1994), we
determined the mean metallicity and age of the stellar population.
Unfortunately, because of the strong emission in $H_\beta$, we were unable
to isolate the absorption line and made estimates by using only metal
absorption lines. Hence we could not reliably separate the effects of a
metallicity and age variations. Therefore, the metallicity near the
photometric center of the galaxy is determined unambiguously, $[Fe / H] =
1$ dex, while the age can range from $12 \cdot 10^9\,M_\odot$ to $17
\cdot 10^9\,M_\odot$ years. At a distance of $4''-5''$ to the NE (in a
region closer to the galactic nucleus), the metallicity and age lie
within the ranges $0.0-0.25$ dex and $1.5-2\cdot 10^9\,M_\odot$ yrs,
respectively. These results provide circumstantial evidence that the
galactic nucleus (where the metallicity must be higher and the age must
be younger) does not coincide with the photometric center.

\subsection{IFP data}

The IFP observations were performed in the $H_\alpha$ and [NII]$\lambda6583$
lines. Since the interferometer has a large field of view, the line-of-sight
velocity fields as well as the distributions of the emission lines and
continuum intensities, lines' $FWHMs$, and lines ratio were constructed for
both components of the interacting system NGC6285/86 (Fig. 5). As we see from
this figure, both the velocity fields and the emission-line images of each
component are peculiar, but the most peculiar features are observed in
NGC6286. Let us consider the data for each galaxy in more detail.

\begin{figure*}
\centerline{\includegraphics[width=15  cm]{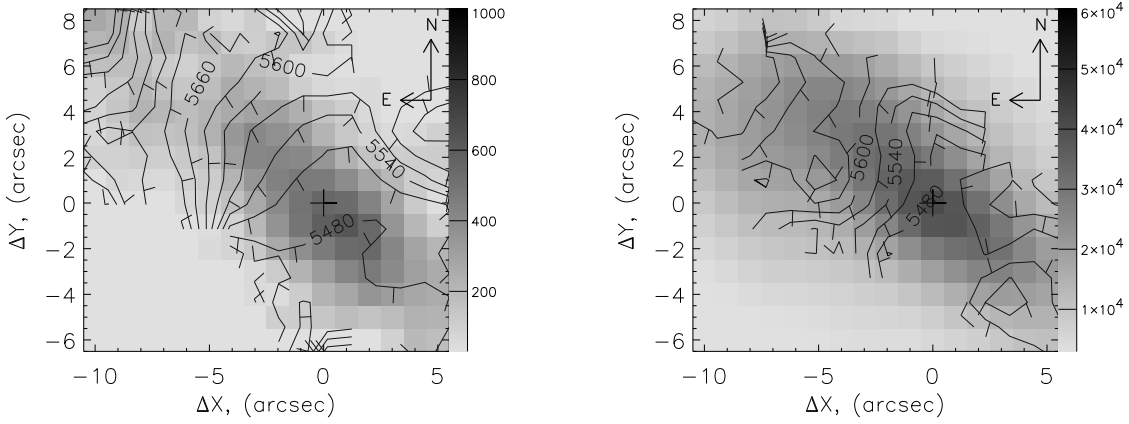}} \caption{ MPFS data.
The line-of-sight velocity fields of the gaseous component found from the
$H_\beta$ (left) and of stars (right) are overlapped on the $H_\beta$ and
continuum intensity distributions (their gray scale is in arbitrary
units). The point with coordinates (0,0) corresponds to the continuum
intensity peak.}
\end{figure*}

\begin{figure*}
\centerline{\includegraphics[width=14  cm,]{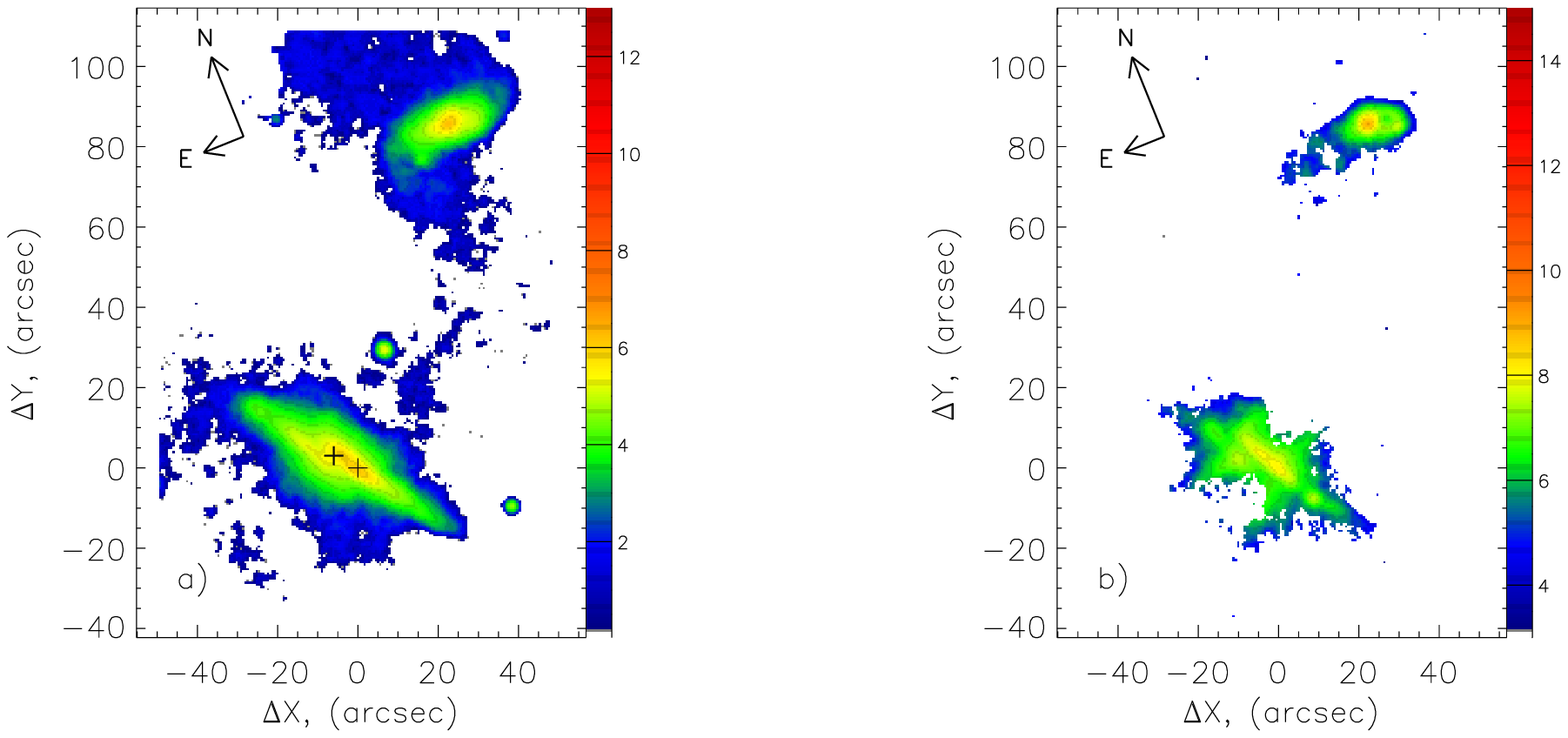}}
\centerline{\includegraphics[width=14  cm,]{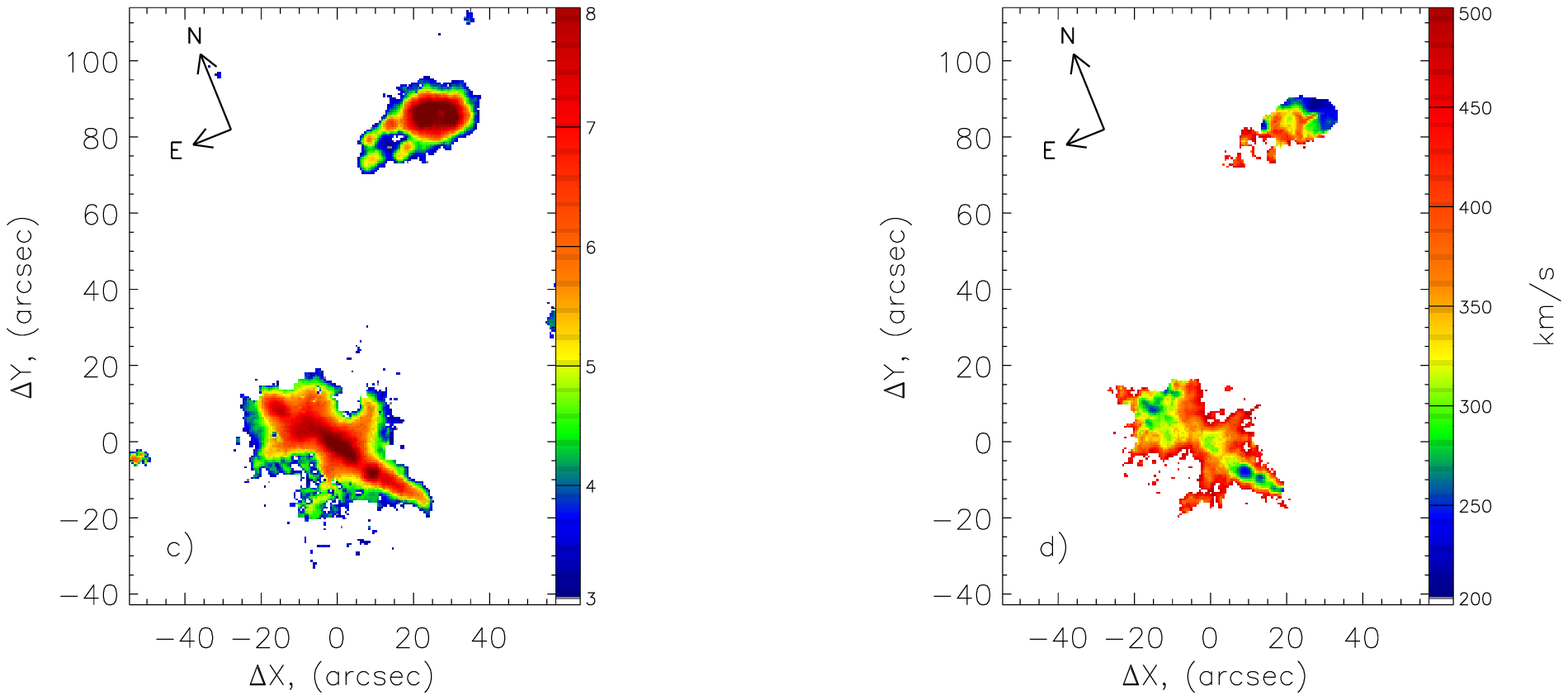}}
\centerline{\includegraphics[width=14  cm,]{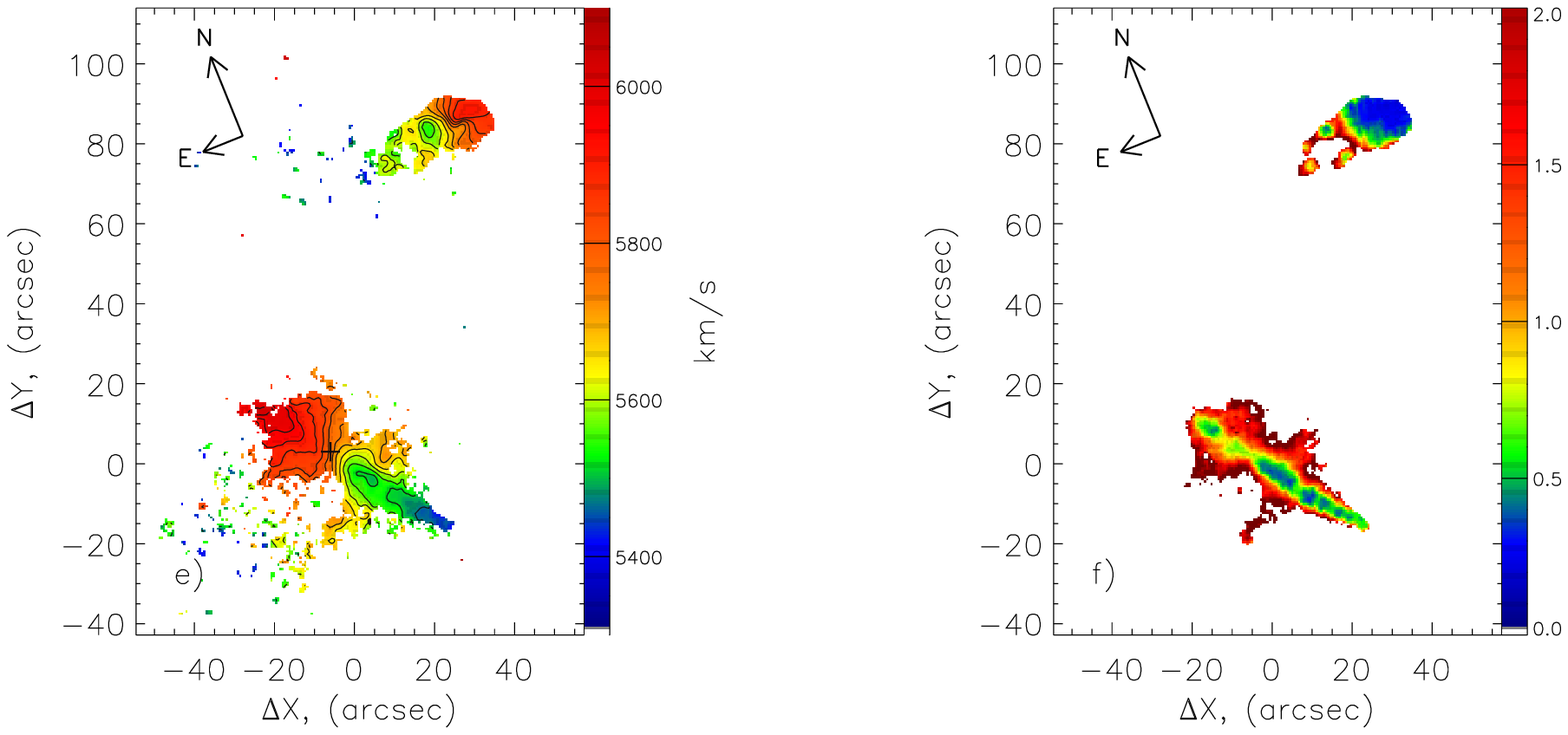}} \caption{  IFP
data: (a) the continuum intensity distribution, the crosses mark the
positions of the photometric $(0,0)$ and dynamical $(6,3)$ centers; (b)
and (c) the intensity distributions in the [NII]$\lambda6583$ and
$H_\alpha$ lines (the gray scale is in arbitrary units); (d) the
distribution of $FWHMs$ for the [NII]$\lambda6583$ line; (e) the velocity
fields in this line, isovelocities are plotted in them at steps of
$40\km$, the cross marks the position of the dynamical center, the
velocity at this point is $5690 \km$; and (f) is the
[NII]$\lambda6583/H_\alpha$ ratio.}
\end{figure*}

\begin{figure}
\centerline{\includegraphics[width=6 cm,angle=-90]{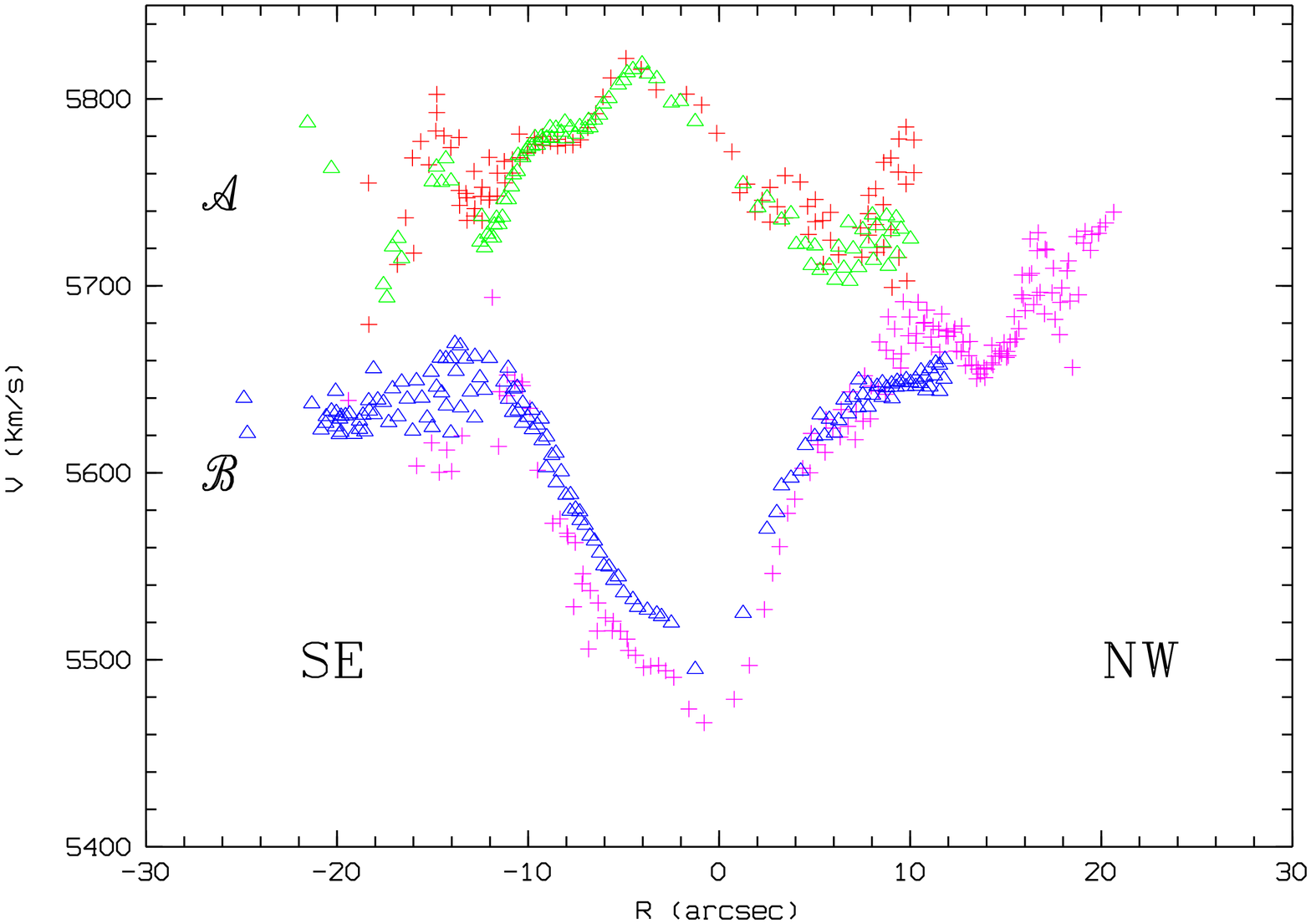}}
\centerline{\includegraphics[width=8  cm]{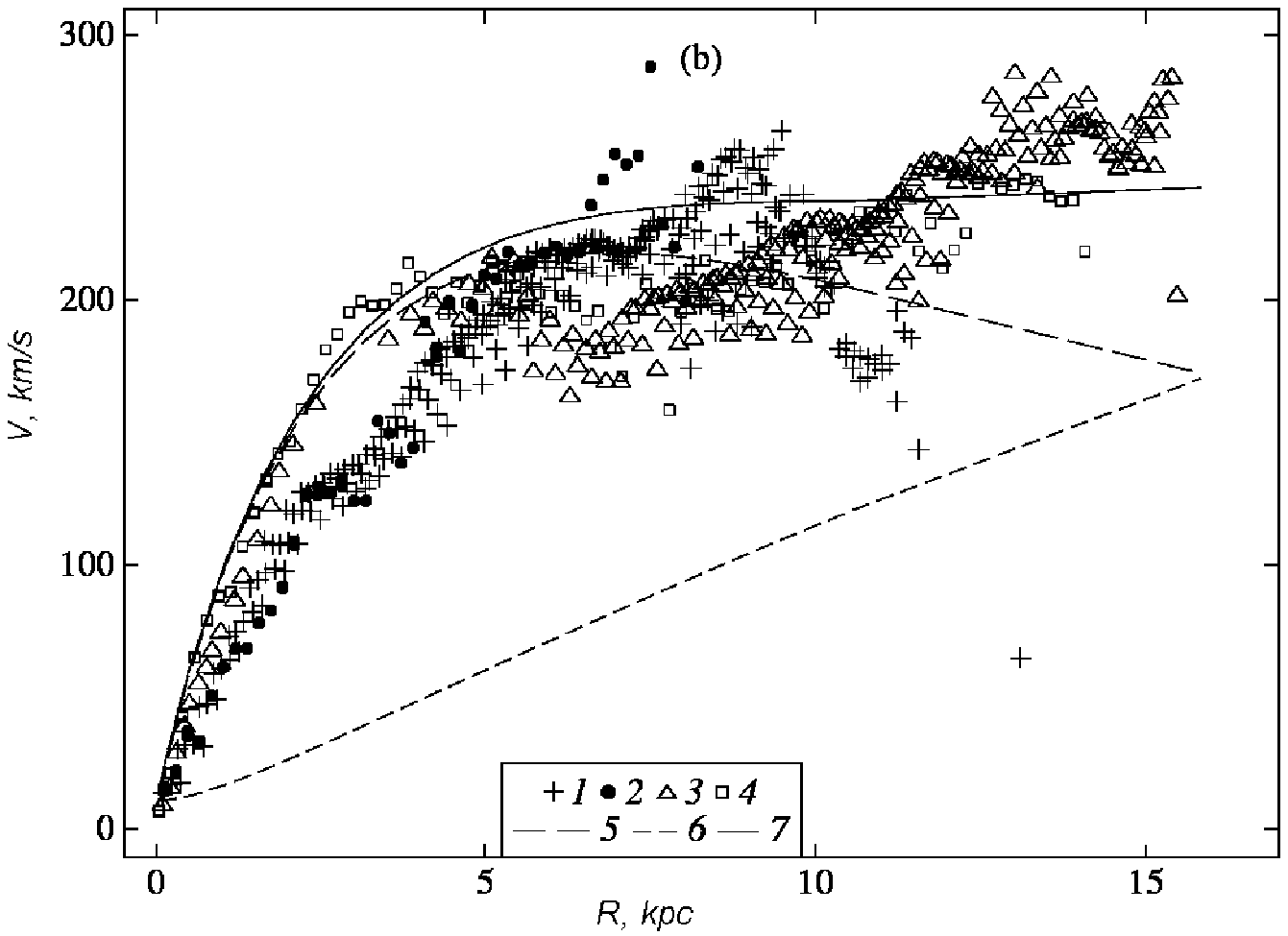}} \caption{(top) The
line-of-sight velocity distributions in the emission-line regions A and B at
slit positions parallel to the minor axis of the galaxy constructed from IFP
observations: the crosses  the $H_\alpha$ data; the triangles  represent the
[NII]$\lambda6583$ data. (bottom) The rotation curve for NGC6286. The NE half
of the line-of-sight velocity curve: 1 and 2 represent the IFP and UAGS data,
respectively; The SW half: 3 and 4 represent the IFP and UAGS data,
respectively; 5 is the model rotation curve for the disk, 6 is the halo model,
and 7 is the ultimate model rotation curve for NGC6286. }
\end{figure}

\subsubsection{NGC 6286}

Intense emission is observed in the central part of the galactic disk,
and a bright H II region can be identified at $\sim 10''$ to the SW from
the photometric center. The [NII]$\lambda6583/H_\alpha$ ratio in the disk
changes slightly; it is approximately equal to 0.35 in the SW part and
is, on average, slightly higher in the NE part, being largest (0.6) near
the dust lane. Since this ratio increases in the dust lane, we may assume
that in this region we see the outer parts of the disk where the physical
conditions differ from the conditions at the center. The relatively small
[NII]$\lambda6583/H_\alpha$ ratio suggests that photoionization is
responsible for the formation of the emission-line spectrum.

A comparison of Figs. 5a, 5b, and 5c shows that the emission-line and
continuum intensity distributions for NGC6286 are different. In all
cases, disk components crossed by the dust lane are traceable. The outer
continuum isophotes are lenticular in shape, while two extended regions
that are symmetric relative to the point shifted from the photometric
center by about $5''-7''$ to the NE in the continuum are observed in the
emission lines. These regions are slightly asymmetric in extent relative
to the disk plane; their SE parts are slightly larger than their NW
parts. Remarkably, with the exception of a few separate knots, there is
no gas emission inside the cones with an opening angle of about
$60^\circ$ the vertices of which are located at the above point. On both
sides of the plane of the galactic disk, the intensity of the [NII] lines
rapidly increases compared to $H_\alpha$ (Fig. 5f). Along the
generatrices of the cones, the line ratio reaches its maximum,
[NII]/$H_\alpha=2-2.5$. Figure 5d shows the distribution of $FWHMs$ for
the [NII]$\lambda6583$ line. The $FWHM$ changes from $\sim200$ to
$\sim320 \km$ in the galactic disk and reaches its maximum, $\sim400
\km$, along the generatrices of the cones.

The emission nebula that we detected in NGC6286 (Figs. 5b and 5c)
resembles in shape the nebulae observed in superwind galaxies, for
example, NGC1482 (Veilleux 2002). Not only the images of both galaxies in
the $H_\alpha$ and [NII] lines but also the distribution maps of the
[NII]$\lambda6583/H_\alpha$ ratio are similar. The shape of the
isovelocities near the gaseous disk is indicative of its rotation around
an axis perpendicular to the galactic plane. However, this axis does not
pass through the photometric center in the continuum, but is shifted by
$5''-7''$ to the NE of it. Its position roughly coincides with the
vertices of the cones in the emission-line images. The extent of the disk
to the SW of the dynamical center is a factor of approximately 1.5 larger
than its extent to the NE, which may be attributable to the disk
asymmetry that resulted from the gravitational interaction with the
companion.

The shape of the isovelocities in emission-line regions outside the galactic
disk (Fig. 5e) is of considerable interest. The velocities are approximately
constant along the straight lines that are parallel to the generatrices of the
cones. For a clearer illustration of the peculiarities of the velocity field,
Fig.~6(top) shows one-dimensional line-of-sight velocity distributions in
directions that are perpendicular to the disk plane and that are offset from
the dynamical center by 2 kpc to the NE (profile `A') and SW (profile `B').
Both profiles pass through the most extended parts of the emission nebula. The
line-of-sight velocity curves are symmetric relative to the disk plane and are
mirror reflections relative to each other. The line-of-sight velocity
amplitude in profile `A' that passes through the dust lane is smaller than
that in profile `B'. The central parts of the profiles reflect the rotation of
the gaseous disk; both profiles flatten out in the outer parts, with the mean
velocities being approximately equal to $5720$ and $5650 \km$, respectively.
As we see from the velocity field, there is no evidence of gas motion around
the major axis of the galaxy.

\subsubsection{NGC 6285}

 The emission-line intensity distribution for the
companion galaxy (Fig. 5b) exhibits the following features: a gaseous
disk, intense emission in the nucleus, and a bright HII region $\sim7''$
to the west of the nucleus. The nucleus and the HII region lie on the
opposite sides of the ring-like structure ($7''$ in diameter) that is
clearly seen in the $H_\alpha$ and [NII] images. The SE side of the
gaseous disk is elongated and bent toward the neighboring galaxy. In the
unperturbed western part of the disk, the [NII]$\lambda6583/H_\alpha$
ratio is almost constant, being $\sim0.4$, which suggests that
photoionization is responsible for the formation of the emission. In the
region where the disk is bent, the ratio changes randomly and, on
average, is approximately equal to 1.3. Peculiar features in the
line-of-sight velocity field can also be seen in this part of the disk
(Fig. 5e). In general, the velocity field of NGC6285 is characteristic of
an inclined gaseous disk with solid-body rotation in its central part.
However, the isovelocities are distorted both at the center and on the
periphery, particularly in the SE part of the galaxy.

\section{ANALYSIS OF OBSERVATIONAL DATA AND DISCUSSION}

\subsection{The Kinematics of Gas in the Galaxies of the Pair}

\subsubsection{NGC 6286}

 Since NGC6286 is seen nearly edge-on ($i = 89^\circ$; Smith et al. 1996),
 we constructed the rotation curve for this  galaxy
from our line-of-sight velocity measurements of the $H_\alpha$ and [NII] lines
along its major axis by using both long-slit spectra and IFP data. The
direction of the rotation axis and the position of the dynamical center were
determined from the velocity field (Fig. 5e). Subsequently, we improved this
position by achieving the closest coincidence in the region where the gradient
of both halves of the line-of-sight velocity curve reached a maximum and where
the latter flattened out. As a result, we obtained $V^{sys}_{hel} = 5690 \pm
10 \km$ for the heliocentric velocity of the galaxy, which yields
$V^{sys}_{gal} = 5925 \km$.

Figure 6(bottom) shows the rotation curve on which the data for the two halves
of the line-of-sight velocity curve are plotted by different symbols. Apart
from the central region, $R \leq 1$ kpc, there is good agreement between the
SW and NE parts of the curve only in some segments. Some differences between
the two parts of the curve can be easily explained. For example, the decrease
in velocity at $1.5 \leq R \leq 4.5$ kpc in the NE part of the curve stems from
the fact that the dust lane passes through this region. Dust  to decrease the
velocity in the rotation curves of disk galaxies seen almost edge-on (Bosma et
al. 1992) . The thickest part of the dust lane ends at a distance $R \sim 5$
kpc, and both branches of the rotation curve converge. Further out, we see a
local minimum of the line-of-sight velocity on the SW branch at $5.5 \leq R
\leq 7.5$ kpc. The giant H~II region mentioned in Section 2.3, which probably
has local non-circular motions, is located precisely in this place. Note that
the waves in the line-of-sight velocity curves caused by  various factors are a
common phenomenon.

The initial part of the rotation curve, up to $R \sim 5$ kpc along the SW
half, is well fitted by the model curve (Monnet \& Simien 1977) that
corresponds to the rotation of an exponential disk with a scale factor $h= 3$
kpc (the curve plotted by long dashes in Fig. 6bottom). However, at a distance
$R \geq 5$ kpc, the points lie above this curve. Therefore, we must assume the
existence of a spherical isothermal halo to represent the observed curve over
its entire length. Calculations show that this halo must have the following
parameters: the central density is $\rho_0 = 0.017\,M_\odot\, pc^3$ and the
core radius is $r_c = 84$ kpc (Fig. 6, short dashes). The ultimate theoretical
rotation curve is represented by the solid line in Fig. 6b. This curve
exhibits a rapid rise up to $R \sim 5$ kpc, which then slows down; the
velocity reaches $V = 240 \km$ at $R \sim8$ kpc, which yields an estimate of
the mass within this radius, $1.1\cdot  10^{11}\,M_\odot$. The total mass of
the galaxy with the halo is $2 \cdot 10^{11}\,M_\odot$.

We particularly emphasize that we have found no spectroscopic evidence for the
existence of a kinematically decoupled system that rotates in the polar plane
with respect to the galactic disk. The line-of-sight velocity distributions
shown in Figs. 5e and 6(top) indicate that in regions far from the disk plane,
the velocities are close to the velocity of the galactic center; i.e., they
correspond to those that might be expected in a spherically symmetric halo.

\subsubsection{NGC 6285}

As we noted above (Section 2.3), the velocity field of the companion
galaxy NGC6285 generally corresponds to the rotation of an inclined
gaseous disk. We fitted our data by the model of gas rotation in circular
orbits with the following parameters: the heliocentric line-of-sight
velocity of the dynamical center is $5670 \km$ ($V^{sys}_{gal} = 5905
\km$) and the inclination of the disk plane to the plane of the sky is
$i\sim60^\circ$. The velocity turned out to reach its maximum, $184 \km$,
at a distance of $\sim 4$ kpc. The mass within the radius of 2 kpc is
$10^{10}\,M_\odot$. Regions with high residual velocities ($\pm 50\km$)
can be identified in the region of the ring-like structure between the
nucleus and the bright H II region and on the SE side of the disk where
the isovelocities are distorted. The high residual velocities are
indicative of appreciable non-circular motions in these regions.

\subsection{The Superwind in NGC6286}

Both the photometric and spectroscopic data suggest that NGC6285/86 is a
pair of interacting galaxies. The strong interaction between them is
evidenced by the distortions of the structure of the galaxies -- the
asymmetry of their gaseous disks and the inclined dust lane in NGC6286;
the existence of a bridge between the galaxies and an extended luminous
region to the SE of NGC6285; and peculiarities of the velocity  fields.
The most probable range of relative space velocities of the galaxies
during their close encounter is $50-100 \km$. The probability that their
relative space velocity is outside this range for the derived
line-of-sight velocity difference between the galaxies of $\sim20 \km$ is
low. We can then estimate the characteristic interaction time, about
$10^8$ yrs.

The interaction during a close passage of the galaxies could trigger a strong
starburst in the central part of NGC6286. This conclusion is confirmed by FIR
and IR data, according to which NGC6285 is a galaxy with a high FIR
luminosity: $\log(L_{FIR}/L_\odot) = 11.28$. Such a luminosity $L_{FIR}$ is
suggestive of a high star-formation rate: $56-65\,M_\odot\,yr^{-1}$ (Smith et
al. 1998; Poggianti \& Wu, 2000).

This a  high rate of star formation, which, in addition, takes place in a
relatively small region in the central part of the galaxy, eventually
leads to a much higher supernova rate than that in normal spiral
galaxies. As a result, a phenomenon called a superwind will be arised
(Heckman et al. 1990). This phenomenon consists in the strong heating of
the interstellar medium and its outflow in a direction perpendicular to
the galactic disk. X-ray emission from the outflowing hot gas (which was
actually detected from such well-known superwind galaxies as M82 (Lehnert
et al. 1999) and NGC~253 (Weaver et al. 2002)) and the presence of an
optical emission nebula serve as observational manifestations of this
phenomenon. In this case, the edge-on orientation of the galaxy is most
favorable for the detection of a superwind.

Let us consider the observational evidence suggesting the existence of a
superwind in NGC6286 in more detail. According to RASS\,FSC data, this
galaxy is an X-ray source. We detected an optical emission nebula in the
galaxy (Figs. 5b and 5c) the size of which reaches $\sim9$ kpc in a
direction perpendicular to the plane of the stellar disk. Note that the
sizes of such nebulae in other superwind galaxies (M~82, NGC~1482, and
others) lie within the range from several kpc to several tens of kpc
(Heckman et al. 1990). As we noted above, no optical emission is observed
inside the cones, because, according to the calculations by Strickland
and Stevens (2000), the gas temperature in this region must be
$\sim10^7$~K.

A good illustration of a superwind in a galaxy is an enhanced
[NII]$6583/H_\alpha$ the emission-lines ratio compared to an ordinary H~II
region. Thus, whereas this ratio is $\leq0.5$ in an H II region where
photoionization dominates, it is much larger than 0.5 for a supersonic
outflow of hot gas and the formation of shock fronts (the conditions under
which collisional ionization dominates). We see from Fig. 5f that the
derived ratios increase severalfold as one recedes from the disk plane.
In this case, the emission-line ratio is highest in the regions that form
the boundary of the conical ejections of hot gas from the galactic plane.

The line emission is strongest where the line of sight runs along the
walls of the cone. The space velocity here is roughly perpendicular to
the line of sight, and the line-of-sight velocities cannot give evidence
of such motions. However, the increase in emission lines' $FWHMs$ to $400
\km$ near the walls of the cone and their complex profiles, as was
mentioned above, serve as their indirect confirmation.

Thus, we conclude that a superwind-type bipolar outflow of gas from the
central region of the disk takes place in the galaxy NGC6286.

We note in passing yet another fact that confirms the interaction between
the galaxies. The [NII]$\lambda6583/H_\alpha$ ratio also increases in the
SE part of the galaxy NGC6285 that is closest to NGC6286 (Fig. 5f). This
increase may be attributable to the hot superwind gas that heats up and
ionizes the interstellar medium in this part of the galaxy, which causes
the intensity of the forbidden lines to increase.

Using the typical superwind outflow velocities ($\sim500 \km$) from
Veilleux (2002) and the observed sizes of the ionization region above and
under the disk, we can estimate that the ejection began $\sim10^7$ years
ago. This estimate is consistent with the above estimate for the
characteristic time of strong interaction.

\section{CONCLUSIONS}

 In conclusion, we emphasize that our kinematics study of
the gaseous component of the galaxy NGC6286 has revealed no evidence of
gas rotation around its major axis. Therefore, NGC6286 is unlikely to be a
galaxy with a forming polar ring, as assumed previously.

Based on the entire set of observational data, we have concluded that the
peculiarities of NGC6286 are attributable to the presence of a superwind
that outflows from its central region. This is suggested by the following
facts:

\begin{itemize}

\item the existence of an emission nebula stretching to a distance of
$\sim9$ kpc from the galactic plane whose shape corresponds to a bipolar
outflow of hot gas; the presence of this outflow is confirmed by the
presence of an X-ray source;

\item the significant increase in [NII]$\lambda6583/H_\alpha$ ratio
characteristic for superwind galaxies;

\item the high infrared luminosity of
the galaxy, which is indicative of a high star-formation rate ($SFR
=56-65\, M_\odot\, yr^1$). The starburst in the central region of the
galaxy that gave rise to the superwind was probably triggered by a close
encounter of the two galaxies.

\end{itemize}

\begin{acknowledgements}

We are grateful to the Large Telescopes Program Committee  for allocating
observational time on the 6m telescope and to E.V. Volkov for
participation in the discussion of our results. This work was supported
in part by the Astronomy Federal Program (project no. 40.022.1.1.1001),
the Russian Foundation for Basic Research (project no. 02- 02-16033), and
the Russian Ministry of Education (project no. E02-11.0-5).

\end{acknowledgements}

{}

\textit{Translated by V. Astakhov}

\end{document}